\documentclass[doublecol]{epl2} 
\usepackage{graphicx}%

\newcommand{\be}{\begin{equation}}
\newcommand{\ee}{\end{equation}}
\newcommand{\bea}{\begin{eqnarray}}
\newcommand{\eea}{\end{eqnarray}}

\title{Enhanced synchronization in an array of spin torque nano oscillators in the presence of oscillating external magnetic field} 
\shorttitle{Enhanced synchronization in an array of STNOs}
\author{B. Subash\inst{1} \and V. K. Chandrasekar\inst{2} \and M. Lakshmanan\inst{1}}
\shortauthor{B. Subash \etal}

\institute{                    
  \inst{1}  Centre for Nonlinear Dynamics, Department of Physics, Bharathidasan Univeristy, Tiruchirapalli - 620 024, India.\\
 \inst{2} Centre for Nonlinear Science \& Engineering, School of Electrical \& Electronics Engineering, SASTRA University, Thanjavur - 613401, India. \\
  
}

\abstract{
We demonstrate that the synchronization of an array of electrically coupled spin torque nano-oscillators (STNO) modelled by Landau-Lifshitz-Gilbert-Slonczewski (LLGS) equation can be enhanced appreciably in the presence of a common external microwave magnetic field. The applied microwave magnetic field stabilizes and enhances the regions of synchronization in the frequency determining parameter space. We have identified two regions of synchronization in the parameter space of our analysis, where the oscillators are exhibiting synchronized oscillations thereby emitting improved microwave power. To characterize the synchronized oscillations we have calculated the locking range in the domain of external source frequency.
}
\begin{document}





\maketitle 

\section{Introduction} 
The structure of an extended giant magnetoresistance trilayer device consists of a thick fixed ferromagnetic layer or pinned layer, a metal spacer (conducting or nonconducting), and a thin free ferromagnetic layer. When an electric current is injected into the fixed layer, the electrons get spin polarized along the direction of magnetization of the fixed layer. The spin polarized electric current exerts a torque on the magnetization of the free layer which leads to a rich variety of dynamics\cite{Laksh:12,Bazaily:98,Yang:07,Murugesh:09b}. This phenomenon was independently predicted by J. C. Slonczewski\cite{Slonczewski:96} and L. Berger\cite{Berger:96}. For a typical ferromagnetic layer to produce such a spin polarized current one requires a high density electric current of the order $10^6 A/cm^2$, which can be achieved through a nano-contact. The torque acting on the free layer due to spin polarized current is universally called the spin transfer torque, which can either switch the magnetization or induce steady-state precession under appropriate conditions. This trilayer nanoscale device capable of converting direct current into magnetization oscillations which in turn gives an intrinsic high frequency varying magneto-resistance and thus a varying current is termed as a spin torque nano oscillator(STNO). Although this nanoscale level of microwave source has several advantages over the conventional device ranging from wide band of frequency tunability, low bias voltage, resistance to external radiation, to easy integration with CMOS circuits through MRAMs, a single important demerit that makes the device incapable of practical implementation is its low power output ($\sim$ nW). To overcome this difficulty one can phase lock a large array of STNOs which oscillates monotonically\cite{7,Li:06,Georges:08,Georges:08a}.

Phase locking of a single STNO of certain archetype depends highly on the spatial symmetry of the external source and hence a uniform microwave magnetic field can synchronize more efficiently than a common microwave current \cite{grollier:br12}. For an array  of electrically decoupled STNOs an applied external microwave magnetic field/dipolar magnetic field generated by an individual oscillator can act as a medium to phase lock each other\cite{epl:13,kaka}. But the latter mechanism has been proved to be effective only for certain number of oscillators due to the limit in the interpillar distances \cite{grollier:apl13, kaka, mancoff:n05, tabor:l10}. It is important to note that in both the mechanisms each STNO should have separate DC power source to make the device electrically uncoupled. For a large number of oscillators say N, it requires N number of DC power sources to electrically decouple the oscillators. Consequently the number of electrical connections in the device gets increased. This particular need makes these mechanisms difficult to implement in a large array of STNOs so as to convert it into a single device of improved power. Further it has been reported that electrical connections between the oscillators increase the number of synchronized attractors\cite{akerman:b12}. So switching between these attractors due to noise makes the device more complex and unstable in the synchronized state which renders the device unable to tune the output frequency. From an application point of view and for the purpose of  device integration with electronic circuits it is essential to enhance the  synchronization of electrically coupled STNOs \cite{slavin:2009,Nakada:12,Turtle:13}. Hence our main aim is to increase the maximum power ouput region in the frequency determining parameter space.

In this paper, we report that synchronization of electrically coupled STNOs  in the frequency determining parameter space can be enhanced appreciably by using an external microwave magnetic field.  The applied oscillating magnetic field synchronizes the phase desynchronized oscillations due to the presence of electrical connections between the oscillators. We have identified two definite regions of synchronization in the parameter space of our analysis where the microwave power output is maximum. We have compared the regions of synchronization due to the presence of oscillating magnetic field with its absence. The time evolution of the magnetization is shown for both the cases to elucidate the effect of oscillating magnetic field. We have determined the frequency and enhanced power output of an array of STNOs by Fourier transforming the microwave output current. We also report that the oscillating magnetic field increases the power of electrically coupled STNOs by 17 times. To characterize the observed synchronization phenomenon we calculate the locking range in the domain of external source frequency.
\begin{figure}
\includegraphics[width=0.48\textwidth]{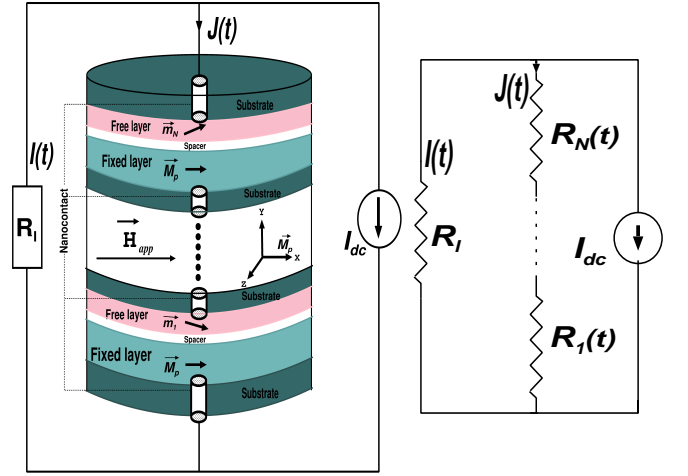}
\begin{center}
\caption{(Color online) Schematic representation of a nanowire consisting of an array of serially coupled STNOs (containing a thick fixed and a thin free ferromagnetic layer separated by a nonmagnetic conducting spacer) connected through nanocontacts biased by a  DC ($I_{dc}$) current source injected into the free layer (taken as negative since electrons are flowing from fixed to free layer). An oscillating external magnetic field ($\vec \mathcal{H}_{app}$) is applied uniformly throughout the array of STNOs. Fixed layer magnetization $\vec M_p$ is along the x axis and it is pinned through a strong anti-ferromagnetic coupling. The synchronized microwave current ($I(t)$) is measured across the load resistance $R_l$ = 20 $\Omega$.}
\label{model}
\end{center}
\end{figure}
\section{Model and description}
Considering a nanowire \cite{dubois:b99,Georges:06}, composed of GMR trilayer structure connected serially through nanocontacts, as shown in Fig. \ref{model}, we study its dynamics. The macrospin magnetodynamics of the $i^{th}$ free layer is described by the LLGS equation\cite{Landau:35,Lakshmanan:84,Hillebrands:02} for the normalized unit magnetization vector $\vec{m}_i = m^x \hat i + m^y \hat j + m^z \hat k$, $|\vec{m}_i|^2=1$. It reads
\begin{eqnarray} \nonumber
\frac{d\vec m_{i}}{dt}  = && -\gamma\vec m_i \times
\vec H_i^{eff}+\alpha\vec m_i \times\frac{d\vec m_i }{dt} \\ 
&&-\gamma \beta(t) \vec m_i \times(\vec m_i \times\hat M_{p}), {\small i=1,2,...N}.
\label{coupled}
\end{eqnarray}
The free layer is a permalloy thin film of dimension $100~nm \times 200~nm \times 5~nm$ extended along the $x-y$ plane. $\hat M_p$ is the normalized magnetization of the fixed layer pinned along the x-axis. The effective field for the $i^{th}$ STNO is given by\begin{equation} 
H_i^{eff} = \vec \mathcal{H}_{app} + \kappa_i m_i^x\hat{i} - 4\pi M_0 m_i^{z}\hat{k},
\label{eff}
\end{equation}
 which comprises of an external applied magnetic field 
 \begin{equation} \vec \mathcal{H}_{app} = (h_{dc}+h_{ac}\cos\omega t)(\sin \vartheta \cos \varphi,\sin \vartheta \sin \varphi, \cos \vartheta),
 \label{app}
\end{equation} a uniaxial anisotropy of strength $\kappa_i$ along the easy axis (x-axis) and the demagnetization field due to the shape of the film  along the easy plane. We choose the parameters involved in Eq. (\ref{coupled}) as the gyromagnetic ratio $\gamma=0.01767~$ Oe$^{-1}$ ns$^{-1}$, the Gilbert damping coefficient $\alpha=0.02$, the density of the spin current $\beta(t)={\hbar \eta J(t)}/{2M_0Ve}$, where $\eta = 0.35$ is the spin polarization ratio, $J(t)$ is the shared microwave current defined in Eq. (\ref{jt}), $M_0$ is the saturation magnetization taken as $4\pi M_0 = 8.4$  {kOe} for a typical bulk permalloy thin film and $V$ is the volume of the free layer.  The inevitable errors in nanolithography process is taken into account by considering the device variability($\Delta\kappa_i /\kappa_1 \times 100\% = 2.22\%$) of STNOs\cite{Persson:07}. The shared microwave current which flows across the STNOs (Fig.\ref{model}) is derived using current divider relation between the serially connected STNOs of resistance $R_i(t)=  R^0_{i}+ \Delta R_i\cos [\theta_i(t)], R^0_{i}=(R^p_i + R^{ap}_i)/2, \Delta R_i=(R^{ap}_i - R^{p}_i)/2 $  and the load resistance $R_l$. It is given as \cite{Georges:06,Persson:07}
\begin{equation}
J(t)= \frac{R_{l} I_{dc}}{R_l + R^0_{i} - \Delta R_i \sum_{i=1}^N \cos \theta_i(t)},
\label{jt}	
\end{equation}
where $R^p_i, R^{ap}_i, \theta_i(t)$ are the resistance, magnetoresistance and the instantaneous angle between free and fixed layers of the $i^{th}$ STNO, respectively. 

\begin{figure}
\includegraphics[width=0.49\textwidth]{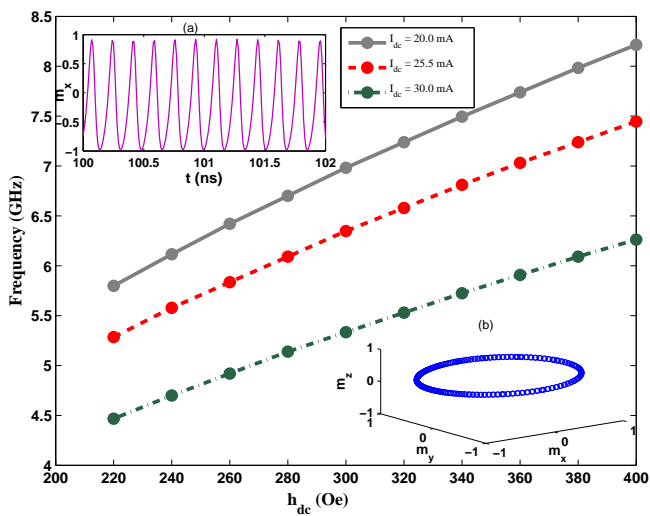}
\begin{center}
\caption{(Color online) Magnetization frequency of a single STNO in the absence of microwave source is plotted for increasing values of static external magnetic field strength($h_{dc}$)  for three different values of the dc current $I_{dc}$. The insets (a) and (b) depict (a) the x-component and (b) the spin $\vec m$ for a specific value of $h_{dc} = 260$ Oe  and $I_{dc} = 25.5 $ mA exhibiting out of plane oscillation.}
\label{self}
\end{center}
\end{figure}
\section{Characteristics of a single STNO}
To start with, we consider the typical nonlinear behaviour of a decoupled STNO in the absence of any external microwave source. In the absence of electrical connection, we have $J(t)=I_{dc}$ in (4). Further, in the absence of external microwave field $h_{ac} = 0$, we choose ($\vartheta$,$\varphi$)=($\pi/2$,0) so that a static field of strength $h_{dc}$ is applied along the x-direction.  For the above choice of parameters, a single uncoupled STNO exhibits limit cycle oscillations in the microwave range. The corresponding  frequency of the single STNO can be tuned by varying the input direct current($I_{dc}$) and/or the static magnetic field strength($h_{dc}$). The corresponding variation of frequency as a function of the strength of the dc field $h_{dc}$ is shown in Fig. \ref{self} for a fixed $I_{dc}=14.13$ mA, which is obtained through a power spectral analysis of the time series $m_x(t)$. Further the time series of $m_x$ and phase portrait of $\vec m$ are plotted as insets for a specific strength $h_{dc} = 260 $ Oe. It may be seen in inset (b) that $\vec m$ is exhibiting an out of plane oscillation which is in confirmation with the fact that the magnetization frequency is increasing with the dc field strength. Similarly, for fixed $h_{dc}$, the frequency of the STNO can be tuned by varying the input bias current $I_{dc}$.

\section{Array of serially connected STNOs under common external microwave magnetic field }
\begin{figure}
\includegraphics[width=1.0\linewidth]{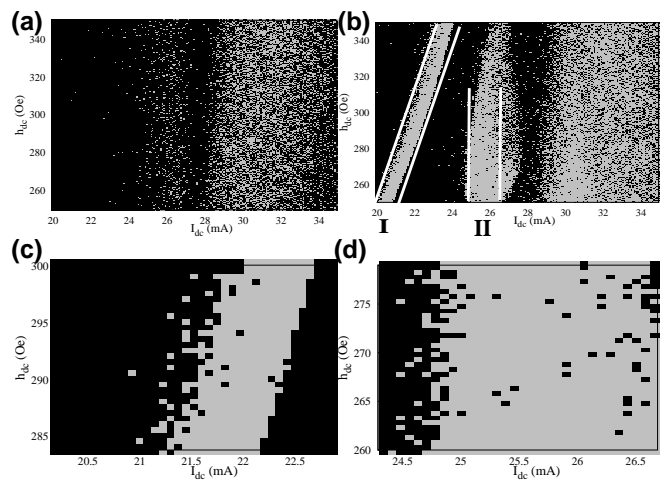}
\begin{center}
\caption{Synchronization dynamics (measured in terms of standard deviation $\sigma$ ) of an array of $N=500$ STNOs for $h_{ac} = 16$ Oe in the frequency determining parameter space, namely input dc current ($I_{dc}$) and dc component of the oscillating magnetic field ($h_{dc}$). The grey region corresponds to the in-phase synchronization of all the STNOs which in turn can deliver the maximum microwave power output, while the black region corresponds to desynchronization dynamics/minimum power output regions. (c) and (d) correspond to enlarged portions in the regions I and II respectively of Fig(b). The oscillation frequencies are 2.392 GHz in region I and 1.196 GHz in region II.}
\label{sigma}
\end{center}
\end{figure}
The macromagnetic simulation of the considered device consisting of $N=500$ STNOs has been carried out by directly integrating Eq. (\ref{coupled}) using a variable step-size Runge-Kutta method. We have carried out the simulation on a 32 node cluster facility compiling with intel fortran compiler of accuracy upto six digits in the normalization of the magnetization vector ($|\vec m_i|^2=1$).  For an array of oscillators it is necessary to find the regions of synchronization in this parameter space($I_{dc}$-$h_{dc}$), where we can tune the synchronized oscillators without affecting the output power. In order to identify the above regions of synchronization we calculate the time averaged standard deviation ($\sigma$) of the x-component of the magnetization vector ($m^x_i$) $\sigma = \left<\sqrt{\frac{1}{N} \sum_i (m_i^x - \mu)^2}\right>_t, ~~~\mu = \frac{1}{N}\sum_i m_i^x$, where $\mu$ is the average time series of the magnetization of N number of STNOs. It is evident that if all the oscillators are completely synchronized $\sigma$ takes a value zero. On the other hand,  it takes a value greater than zero if the oscillators are desynchronized. The device variability of 2.22\% among the STNOs is considered in terms of uniaxial anisotropy ($\vec H_k = \kappa_i m_i^x\hat{i}$) by randomly distributing the values of $\kappa_i$ between $45 $ to $46$ Oe. We also assume the resistance($R_i^p$) and the magnetoresistance($R_i^{ap}$) of all the oscillators to be 10$\Omega$ and 11$\Omega$, respectively, see Eq. (\ref{app}).

In Fig. \ref{sigma}(a) we have plotted the regions of phase synchronization (grey colour) for the above array of $N=500$ with $h_{ac}=0$ Oe (see Eq. (\ref{eff})), electrically connected STNOs placed in the static external magnetic field applied along ($\vartheta$,$\varphi$)=($\pi/2$,0) emitting maximum power. The random choice of initial values for the magnetization vector ($\vec m_i$) for each point in the parameter space chosen with the constraint ($|\vec m_i|^2=1$) induces switching between multiple synchronization attractors\cite{akerman:b12} existing due to electrical connections. In addition to the randomness in the initial values of magnetization, we have added a small additive noise in Eq.(1) to take care of the effects due to temperature. The lack of definite regimes of synchronization in the parameter space is due to the consequence of this switching between the attractors. Further it is to be noted that there exists maximum power output in several points and also in some small regions in the parameter space which will not help to obtain a consistent practical level microwave power output from the device. Any change in the external parameters such as $I_{dc}$, $h_{dc}$, etc., due to noise/temperature effects, may affect the onset of synchronization in the system and hence affect the output power. In order to overcome this difficulty we proceed as follows. We now switch on the alternating magnetic field  of frequency $\omega=15$ GHz (which is chosen as almost twice the free-running frequency of the single STNO such that it will not affect the measurement of output signal\cite{tabor:l10}), applied along the x-direction and have identified that beyond a critical value  $h_{ac}=16$ Oe  in the range $h_{ac} \in (15.0,20.0)$ Oe the oscillators exhibit synchronized oscillations. We choose for our further study the specific value $h_{ac}=16$ Oe in the following. Note here that the strength of the field is nominal compared to the microwave magnetic field produced so far\cite{tabor:l10}. 
\begin{figure}
\includegraphics[width=0.49\textwidth]{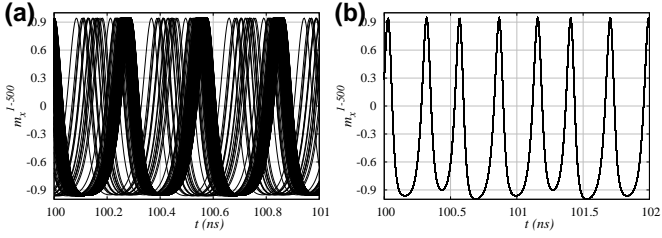}
\begin{center}
\caption{Magnetization dynamics of an array of 500 STNOs, whose device variability is 2.22\%. (a) absence of ac component of external magnetic field ($h_{ac}$=0 Oe) exhibiting phase desynchronized oscillation ($\sigma$ = 0.21), (b) presence of ac component of external magnetic field of strength $h_{ac}$= 16 Oe of frequency $\Omega$ = 15 GHz, exhibiting a synchronized oscillation ($\sigma$ = 0.002). The other parameters $h_{dc} = 251$ Oe, and $I_{dc} = 25.5 $ mA are chosen corresponding to a point in the region II of Fig. \ref{sigma}. }
\label{time}
\end{center}
\end{figure}
The enhanced regions of synchronization (grey colour) in the parameter space due to the above oscillating magnetic field are plotted in Fig. \ref{sigma}(b). We have identified two definite regions(I, II) of phase synchronization in the parameter space where the oscillators are emitting maximum microwave power. Region I (linear diagonal) corresponds to the  maximum power output parameter region where both the parameters can be varied without affecting the output power. On the other hand in region II, which is a vertical region in the parameter space, the input direct current is constant and the static component of the external magnetic field is the varying parameter. The frequency of the output microwave current $I(t)$ in the two regions is calculated as 2.392 GHz in region I and 1.196 GHz in region II. Note that the frequency of the output current is in rational relation with the external source frequency ($\omega=15$ GHz), which is in confirmation with the already reported fractional locking of a single STNO with the external microwave field \cite{tabor:l10}. We have also checked that the phenomenon persists even with larger device variability upto 5\%.
\begin{figure}
\includegraphics[width=0.48\textwidth]{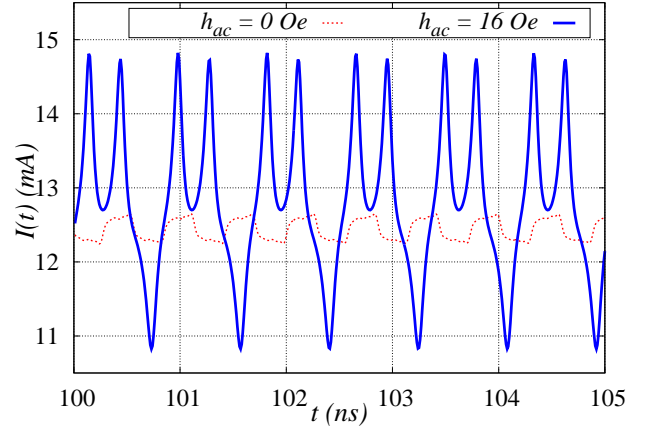}
\begin{center}
\caption{(Colour online) Synchronized microwave current of the array of STNOs measured across the active resistor $R_l$ is plotted for both the cases of absence and  presence of ac component($h_{ac}$) of external magnetic field $\vec \mathcal{H}_{app}$. The other parameters are chosen same as in Fig. 4}
\label{curr}
\end{center}
\end{figure}
The effect of common oscillating magnetic field in the dynamics of x-component of the magnetization vector ($m^x_{1-500}$) of the system of $N=500$ oscillators is shown in Fig.\ref{time} for a point in the region II of Fig.\ref{sigma}(b) corresponding to the values $h_{dc}=251$ Oe and $I_{dc}=25.5 $ mA. The phase desynchronized oscillation of the magnetization in the absence of oscillating magnetic field as in Fig. \ref{time}(a) gets synchronized by the oscillating magnetic field which is clearly elucidated in Fig. \ref{time}(b). We have only presented the x-component of the magnetization vector ($\vec m_i$) exhibiting in-phase synchronization, while the other two components exhibiting anti-phase synchronization dynamics are not presented here because they do not alter the resistance to  the spin current which is flowing along the y-direction with polarization along the x-direction as shown in Fig. \ref{model}, for details see \cite{epl:13}.
\begin{figure}
\includegraphics[width=0.48\textwidth]{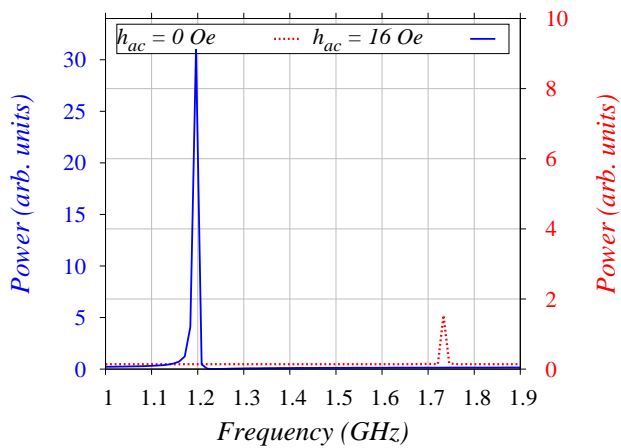}
\begin{center}
\caption{(Colour online) Power spectral density of the microwave current output of an array of 500 STNOs is shown for both the cases of (a) absence of ac component of external magnetic field $\vec \mathcal{H}_{app}$ whose measure is scaled in the right y-axis and  (b) presence of ac component of external magnetic field $\vec \mathcal{H}_{app}$ whose measure is scaled in the left y-axis for the same set of parameters as in Fig. \ref{time}. }
\label{power}
\end{center}
\end{figure}

Next, in Fig. \ref{curr} we have plotted the synchronized  microwave current output $I(t)$ from the array of oscillators,
\begin{equation}
I(t)= \frac{ I_{dc}[R^0_{i} - \Delta R_i \sum_{i=1}^N \cos \theta_i(t)]}{R_l + R^0_{i} - \Delta R_i \sum_{i=1}^N \cos \theta_i(t)},
\label{jt}	
\end{equation} measured across the active load resistor $R_l$. The dotted red line corresponds to the current output in the presence of static external magnetic field of strength $h_{dc}=251.0$ Oe, input direct current $I_{dc}=25.5$ mA and in the absence of oscillating component ($h_{ac}=0$ Oe) of the external magnetic field . The solid blue line corresponds to the output current in the presence of oscillating magnetic field of strength $h_{ac}=16$ Oe and frequency $\omega = 15$ GHz. It is evident that the amplitude of the output microwave current in the presence of $h_{ac}$ is much higher than that in its absence.
\begin{figure}
\includegraphics[width=0.4912\textwidth]{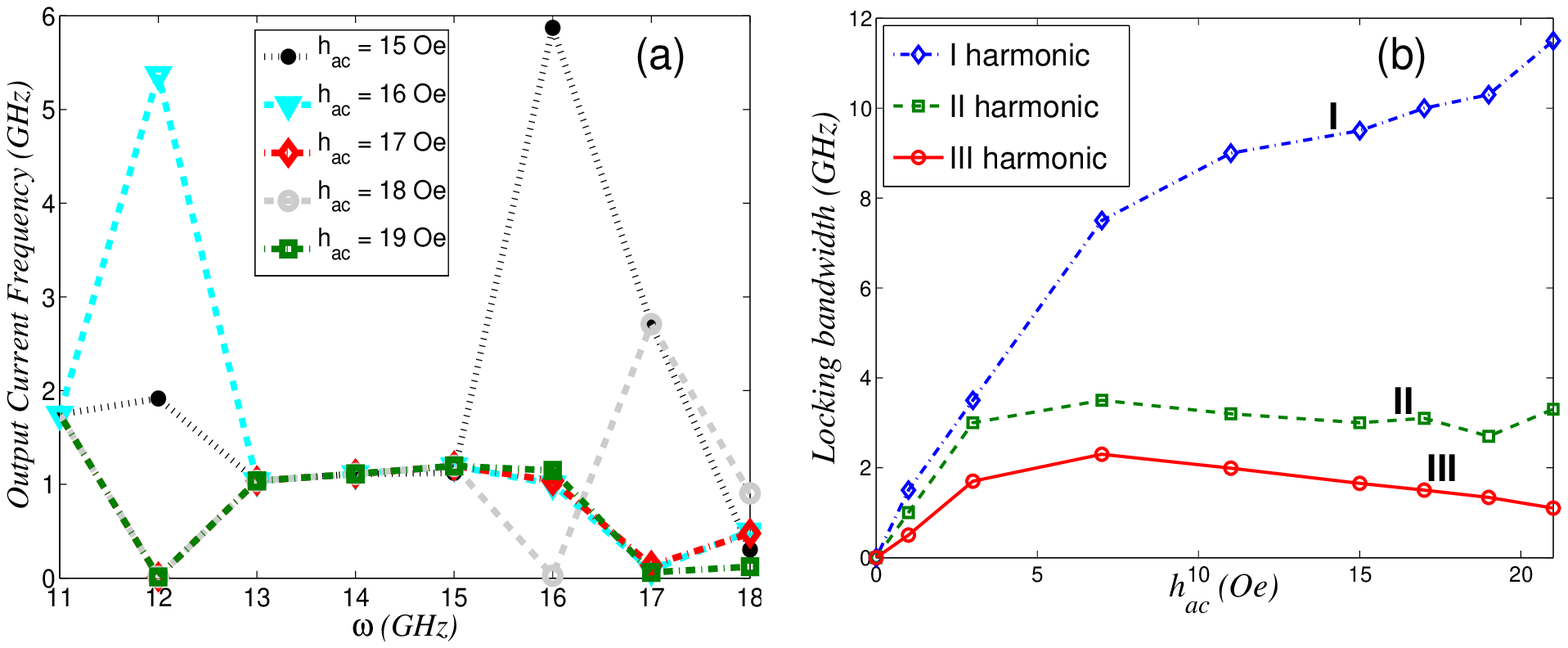}
\begin{center}
\caption{(Color online) (a)Locking range of the synchronized microwave oscillation (output current frequency vs external source frequency) and (b) simplified Arnold tongue diagram computed for different values of amplitude of the microwave magnetic field $h_{ac}$. The other parameters are chosen as in Fig. 4.}
\label{lock}
\end{center}
\end{figure}
In order to elucidate the experimental consequences of synchronization induced due to the oscillating magnetic field, we have plotted the emitted power in the frequency domain by Fast Fourier Transforming the output current $J(t)$ measured across $R_l$ as shown in Fig. \ref{power}. It can be seen that in the presence of $h_{ac}$ the power output (solid blue line) of the STNOs is almost 17 times larger than that in its absence (dotted red line). In Fig. \ref{power}, one may also observe that the linewidth(full width at half maximum) got reduced with the application of oscillating external magnetic field compared to the case of $h_{ac}=0$

To analyze the quality of synchronization we have calculated the locking range in the synchronization region. To elucidate the robustness of the synchronization state observed in Fig. 4, we have varied the frequency of the external microwave field around $\omega = 15$ GHz and calculated the output current frequency through power spectral analysis. Fig. \ref{lock}(a) shows the locking range of the synchronization where it can be observed that the synchronization state withholds  over a range of $h_{ac}$, that is the frequency of the output current remains constant in the external frequency range $\omega = $ 13 to 16 GHz. Any noise in the frequency of the external source in the above range will not affect the synchronization of the serially connected STNOs.  In addition to the fractional locking we have also observed integer harmonic locking\cite{Fino:10} of magnetization precession with the external microwave field. In Fig. \ref{lock}(b), we show the effect of external microwave field amplitude on the locking of STNOs by plotting Arnold tongue diagram corresponding to the locking bandwidth for integer harmonics (I, II and III).  
\begin{figure}
\includegraphics[width=0.49\textwidth]{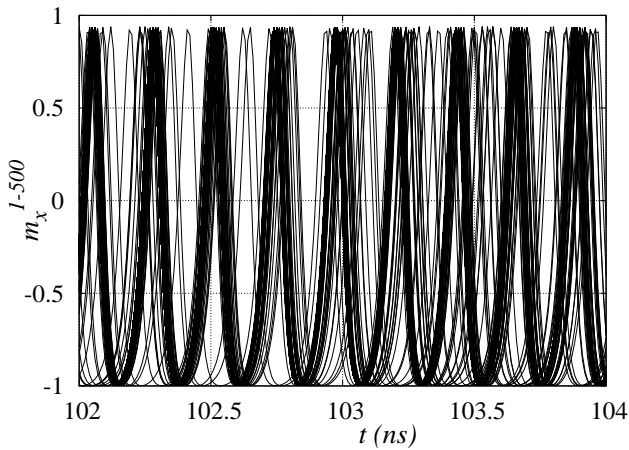}
\begin{center}
\caption{Magnetization dynamics of an array of 500 serially connected STNOs with a common microwave current injection of strength $I_{ac}= 2.55 $ mA ($\omega = 15 $ GHz), ten times smaller than $I_{dc} = 25.5$ mA exhibiting phase desynchronized oscillation. Here the synchronization measure is calculated as $\sigma$  = 0.22.}
\label{iac}
\end{center}
\end{figure}
Note in Fig. \ref{time}(b), we have applied a common microwave magnetic field of strength $h_{ac} = h_{dc}/15.68 = 16 Oe $, ($h_{dc} = 251 $ Oe) which is almost 16 times smaller than the dc component of the external magnetic field. Finally, we also made a comparison of the procedure of applying a common microwave field over that of a microwave current. For this purpose we inject a microwave current of strength $I_{ac}$ along with the DC component $I_{dc}$ so that $I_{dc}$ in Eq.(5) is replaced by the total current ($I_{dc} +I_{ac}\cos \omega t$). We chose the frequency of microwave current $\omega$ the same as that of the field and switched off the ac field $h_{ac} = 0$. The equation of motion for the above configuration is same as (1) with the above changes in parameters. We apply a microwave current of strength $I_{ac} = I_{dc}/10 = 25.5/10 = 2.5$ mA.
The magnetization dynamics ($m_x$) of all the 500 STNOs are plotted in the asymptotic time interval. The dynamics remains phase desynchronized as shown in Fig.\ref{iac} which is further evidenced by the synchronization measure ($\sigma =0.22$). Note that the applied current is now ten times smaller than the direct current, but even this is insufficient to drive the STNOs to a synchronized state.  We also note that the minimum threshold current required for synchronization is $I_{ac}=$ 5.1 mA, that is 5 times that of the dc current. This clearly shows the efficacy of the present method of synchronization using a common external microwave field. 

\section{Summary and conclusion}

In summary, we have demonstrated the significant enhancing effect of an oscillating external magnetic field in the synchronization of an array of serially connected STNOs through macromagnetic simulation of the free layer magnetization vector. In general, specific regions of synchronization have been enhanced due to the presence of oscillating magnetic field. We have identified two definite regions of synchronization in the parameter space where the STNOs are emitting the maximum output power. We have compared our results with the case of microwave current injection of strength comparable with the field and found the strength is insufficient to induce synchronization among such a large array of STNOs. Hence a common oscillating external magnetic field is more efficient than the microwave current injection in case of serially connected STNOs. Further, we wish to emphasize that the considered system with constant frequency external microwave magnetic field can be a vital source of microwave power in telecommunication device tunable over a wide range of frequency with practical level of output power. 
 \acknowledgments
The work forms part of a Department of Science and Technology (DST), Government of India, IRHPA project and is also supported by a DST Ramanna Fellowship of M. L. He has also been financially supported by a DAE Raja Ramanna Fellowship.



\end{document}